\documentclass[aps,showpacs,showkeys,prl,preprintnumbers,nofootinbib]{revtex4}

\usepackage{color}
\usepackage{graphicx}
\usepackage{bbm}

\newcommand{\nc}{\newcommand}
\nc{\non}{\nonumber}
\nc{\hc}{\hbox {H.c.}} 
\nc{\noi}{\noindent}
\nc{\barx}{\bar{x}}
\nc{\pbarn}{\;\hbox {pb}}
\nc{\fbarn}{\;\hbox {fb}}
\nc{\lsp}{\;\;\;\;\;}
\nc{\Lsp}{\;\;\;\;\;\;\;\;\;\;}  
\nc{\LLsp}{\lspace \lspace}
\nc{\lra}{\longrightarrow}
\nc{\beq}{\begin{equation}}  \nc{\eeq}{\end{equation}}
\nc{\bea}{\begin{eqnarray}}  \nc{\eea}{\end{eqnarray}}
\nc{\baa}{\begin{array}}     \nc{\eaa}{\end{array}}
\nc{\bit}{\begin{itemize}}   \nc{\eit}{\end{itemize}}
\nc{\ben}{\begin{enumerate}} \nc{\een}{\end{enumerate}}
\nc{\bce}{\begin{center}}    \nc{\ece}{\end{center}}
\nc{\bpm}{\begin{pmatrix}}   \nc{\epm}{\end{pmatrix}}
\nc{\bvt}{\begin{verbatim}}  \nc{\evt}{\end{verbatim}}

\def\gesim{\,{\raise-3pt\hbox{$\sim$}}\!\!\!\!\!{\raise2pt\hbox{$>$}}\,}
\def\lesim{\,{\raise-3pt\hbox{$\sim$}}\!\!\!\!\!{\raise2pt\hbox{$<$}}\,}

\def\gev{\;\hbox{GeV}}
\def\tev{\;\hbox{TeV}}

\def\m{\;\hbox{m}}

\def\lsp{\qquad}
\def\lsim{\lesim}
\def\gsim{\gesim}
\def\hc{\hbox{H.c.}}

\def\vev{vacuum expectation value}
\def\lcal{{\cal L}}
\def\ocal{{\cal O}}

\def\mati{{\mathbbm1}}

\def\vevof#1{\left\langle#1\right\rangle}

\nc{\Lam}{\Lambda}
\nc{\Lams}{\Lambda^2}
\nc{\mhs}{m_h^2}
\nc{\mws}{m_W^2}
\nc{\mzs}{m_Z^2}
\nc{\mts}{m_t^2}
\nc{\mh}{m_h}
\nc{\mw}{m_W}
\nc{\mz}{m_Z}
\nc{\mt}{m_t}
\nc{\vp}{\varphi}
\nc{\mpl}{m_{Pl}}

\nc{\lamp}{\lambda_H}
\nc{\lamvp}{\lambda_\varphi}
\nc{\lamx}{\lambda_x}
\nc{\xf}{x_f}

\begin{document}

\preprint{IFT-09-01 \cr UCRHEP-T463}

\title{Pragmatic approach to the little hierarchy problem \\
{\it - the case for Dark Matter and neutrino physics -}}

\author{Bohdan GRZADKOWSKI}
\email{bohdan.grzadkowski@fuw.edu.pl}
\affiliation{Institute of Theoretical Physics,  University of Warsaw,
Ho\.za 69, PL-00-681 Warsaw, Poland}

\author{Jos\'e WUDKA}
\email{jose.wudka@ucr.edu}
\affiliation{Department of Physics and Astronomy, University of California,
Riverside CA 92521-0413, USA}

\begin{abstract}

We show that the addition of real scalars (gauge singlets) to
the Standard Model can both ameliorate the little hierarchy problem
and provide realistic Dark Matter candidates.  To this end, the
coupling of the new scalars to the standard Higgs boson must be
relatively strong and their
mass should be in the $1-3\tev$ range, while the lowest cutoff of
the (unspecified) UV completion must be $ \gesim 5\tev$, depending on the
Higgs boson mass and the number of singlets present. The existence of the singlets also leads to
realistic and surprisingly reach neutrino physics.  The resulting
light neutrino mass spectrum and mixing angles are consistent with the
constraints from the neutrino oscillations.

\end{abstract}


\maketitle

\paragraph{Introduction}

The goal of this project is to provide the most economic extension of
the Standard Model (SM) for which the little hierarchy problem is
ameliorated while retaining all the successes of the SM.  
We focus here on leading corrections to the SM, so we will consider
only those extensions that  interact with the SM through
renormalizable interactions (below we will comment on
the effects of higher-dimensional interactions). 
Since we
concentrate on taming the quadratic divergence of the Higgs boson mass, it
is natural to consider extensions of the scalar sector: when adding a new
field $\vp$, the gauge-invariant coupling $ | \vp|^2 H^\dagger H
$ (where $H$ denotes the SM scalar doublet) will generate additional
radiative corrections to the Higgs boson mass that can serve to soften the
little hierarchy problem. 
In this paper we will consider a class of modest extensions 
by adding several real scalar fields 
which are neutral under the SM gauge group.  
The extension we consider, although renormalizable, will
still be understood to constitute an effective low-energy theory valid
up to energies $\sim 5-10 \tev$; we shall not discuss the UV
completion of this model.

\paragraph{The little hierarchy problem}
Within the SM the quadratically divergent 1-loop correction to the
Higgs boson ($h$) mass is given by
\beq
\delta^{(SM)} \mhs =\left[
3\mts/2-(6\mws+3\mzs)/8  - 3\mhs/8  \right]  \Lams/(\pi^2 v^2)
\label{hcor}
\eeq
where $ \Lam$ is a UV cutoff (we use a cutoff regularization)
and $v \simeq 246 \gev $ denotes the \vev\ of the scalar doublet
(SM logarithmic corrections are small since we assume $v \ll \Lam \lesim
10 \tev$); the SM is treated here as an effective theory valid below
the physical cutoff $\Lam$, the scale at which new
physics becomes manifest.

Since precision measurements (mainly from the oblique $T_{\rm obl}$
parameter ~\cite{Amsler:2008zz}) require a light Higgs boson, $m_h
\sim 120-170 \gev$, the correction (\ref{hcor}) exceeds the mass
itself even for small values of $ \Lam $, e.g. for $\mh = 130 \gev$ we
obtain $\delta^{(SM)} \mh^2 \simeq \mh^2$ already for $\Lam \simeq 580
\gev$. On the other hand constraints on the scale of new physics that
emerge from analysis of operators of dim 6 require $\Lam \gsim$ few
TeV.  This difficulty is known as the little hierarchy problem.

There are  two ways to solve this problem: one adds new particles
whose effects either {\it(i)} generate radiative corrections that
partially cancel (\ref{hcor}), as is done in supersymmetric theories
(for which $\delta \mhs \ll \mhs$ up to the GUT scale); or {\it(ii)}
increase the allowed value of $ \mh $ by canceling the contributions
to $T_{\rm obl}$ from a heavy Higgs (see e.g.  \cite{Barbieri:2006dq}).

Here we follow the first strategy, but with a modest goal:
we construct a simple modification of the SM within which $\delta
\mhs$ (the total correction to the SM Higgs boson mass squared) is
suppressed only up to $ \Lam \lsim 3-10 \tev$.  Since (\ref{hcor}) is
dominated by the fermionic (top) terms, the most economic way
of achieving this is by introducing new scalars $\vp_i$ whose
1-loop contributions balances the ones derived from the SM.  In order
not to spoil the SM predictions we assume that $\vp_i$ are singlets
under the SM gauge group. It is then easy to see that the oblique
parameters will remain unchanged if $\langle \vp_i \rangle =0$ (which we
assume hereafter), so that the SM prediction of a light Higgs is
preserved. An extension of the SM by an extra scalar singlet was
also discussed in~\cite{Meissner:2006zh}, there however (classical)
conformal symmetry was adopted to cope with the hierarchy problem.

The most general scalar potential consistent with $Z_2^{(i)}$ independent
symmetries $\vp_i\to -\vp_i$ (imposed in order to 
prevent $\vp_i \to hh$ decays) reads:
\beq
V(H,\vp_i)=-\mu_H^2 |H|^2 + \lamp |H|^4
+ \sum_{i=1}^{N_\vp}(\mu_\vp^{(i)})^2 \vp_i^2 + 
\frac{1}{24}  \sum_{i,j=1}^{N_\vp} \lambda_\vp^{(ij)} \vp_i^2 \vp_j^2
+ |H|^2 \sum_{i=1}^{N_\vp}\lamx^{(i)}  \vp_i^2
\label{pot}
\eeq
In the following numerical computations we assume for simplicity
that $\mu_\vp^{(i)}=\mu_\vp$, $\lambda_\vp^{(ij)}=\lambda_\vp$ and
$\lamx^{(i)}=\lamx$,
in which case (\ref{pot}) has an $ O(N_\vp)$ symmetry
(small deviations from  this assumption do 
not change our results qualitatively).
The minimum of $V$ is at $\vevof H = v/\sqrt{2}$ and
$\vevof{\vp_i} = 0$ when $ \mu_\vp^2 > 0$ and $\lamx, \lamp > 0$ 
which we now assume. The masses
for the SM Higgs boson and the new scalar singlets are 
$\mhs=2\mu_H^2$ and $m^2
= 2\mu_\vp^2+\lamx v^2$ ($\lamp v^2=\mu_H^2$), respectively.

Stability (positivity) of the potential at large field strengths
requires $ \lamp \lambda_\vp > 6 \lamx^2 $ at tree level.
The high energy unitarity behavior (known~\cite{Cynolter:2004cq} for $N_\vp=1$) 
implies $\lamp \leq 4\pi/3$ (the SM requirement) and $\lambda_\vp \leq 8 \pi$, $\lamx <
4\pi$.
Note however that these conditions are derived from the
behavior of the theory at energies $E \gg m$, where we don't pretend
our model to be valid, so that neither the stability limit nor the
unitarity constraints are applicable within our pragmatic
strategy~\footnote{
These conclusions remain even if one includes 
higher-dimensional operators since such terms are
subdominant unless the energies and/or field strengths
are of order $ \Lambda $ -- were the model is
not valid; such operators can also generate
spurious minima, but  these
have scale $ \sim \Lambda $ and are not within the range of validity 
of the model.
It is  also fair to note that for $N_\vp=1$ the stability limit
for $\mh > 115\gev$ implies $\lambda_\vp > 12 (\lamx v/\mh)^2 \gsim 55
\lamx$. Then using $\lambda_\vp \leq 8 \pi$ we find
$\lamx \lesim 0.68$; this does not allow for a significant
cancellation of the SM contributions (\ref{hcor}) and the little
hierarchy problem remains. Increasing $N_\vp$ suppresses $\lamx$
and 
relaxes the unitarity constrains.} that aims at a modest increase of $ \Lam$
to the $3-10\tev$ range.

\begin{figure}[h]
\centering \includegraphics[width=5cm]{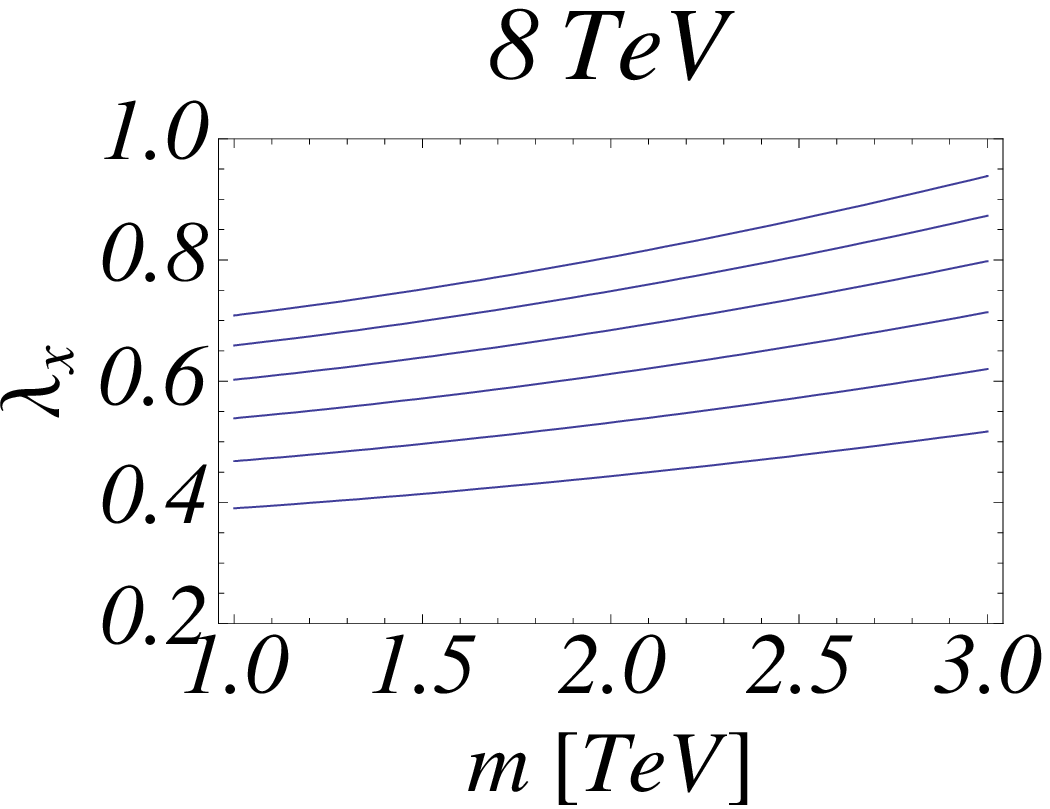}
\centering \includegraphics[width=5cm]{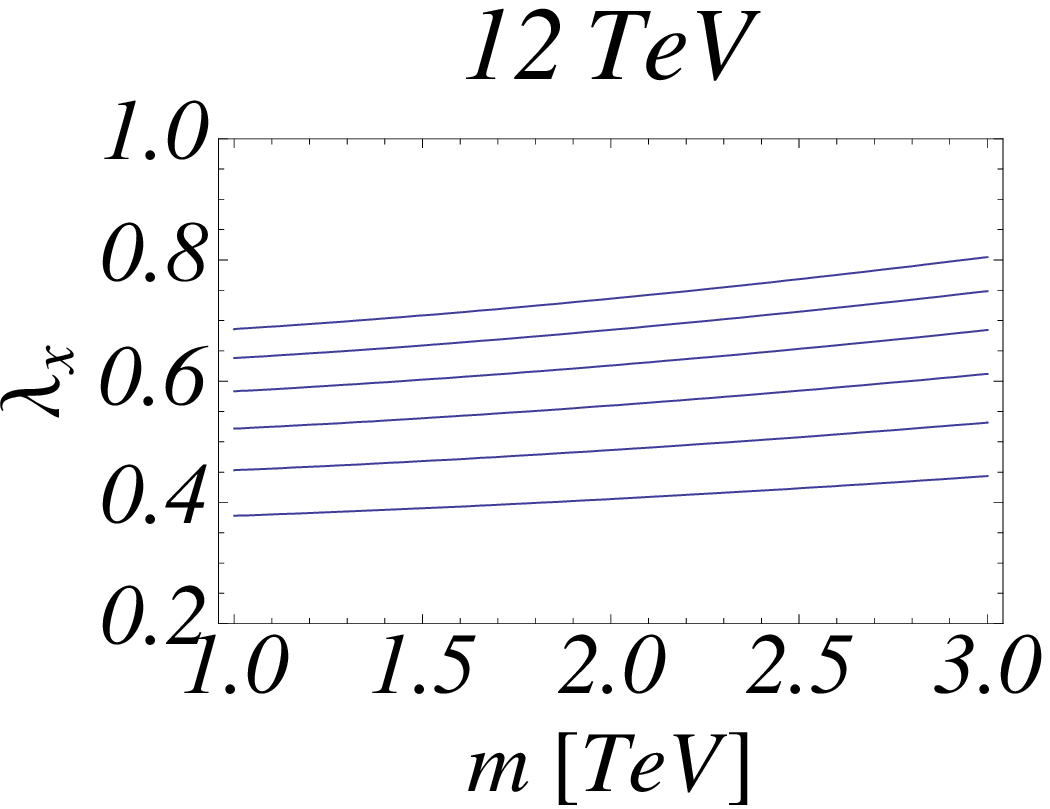}
\centering \includegraphics[width=5cm]{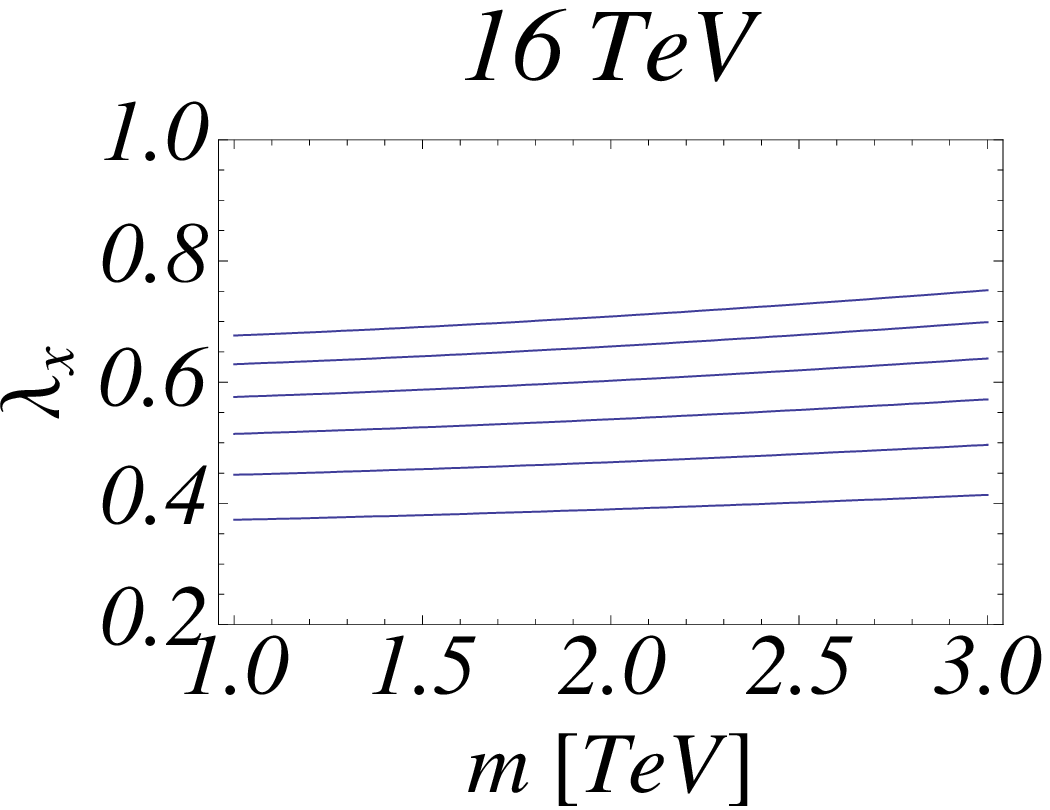}
\caption{Plot of $\lamx$ 
corresponding to $D_t=0$ and $N_\vp=6$
as a function of $m$ for
$\Lam=8,12,16\tev$ (as indicated above 
each panel). 
The various curves correspond to
$\mh=130, ~150, ~170, ~190, ~210, ~230 \gev$ (starting with
the uppermost curve).
}
\label{lam}
\end{figure}

The presence of $ \vp_i $ generates additional radiative
corrections~\footnote{ The $ \Lam^2 $
corrections to $m^2$ can also be tamed within the full model
with additiona fine tuning,
but we will not consider them here, see~\cite{BJ}. The advantage of
the present  model is that this  tuning allows small
corrections to $ m,~m_H $ while keeping them below the cutoff.}
to $\mhs$. However
different ways of imposing the cutoff $\Lam$ (cutoff regularization,
higher-derivative regulators, Pauli-Villars regulators, etc.) yield
different expressions for the extra corrections;  for
large $\Lam$, the coefficients of the $ \Lam^2 $ and $ m^2
\ln(\Lam^2/m^2)$ terms are universal, but but the sub-leading, terms are not. 
Since the sub-leading contributions are small for $m \ll \Lam$ (this
is the range interesting for us) the differences
between various regularization schemes are not relevant. Here
we decided to adopt the simple UV cutoff regularization. Then the extra
contribution to $\mhs$ read
\beq
\delta^{(\vp)} \mhs = - [N_\vp\lamx/(8\pi^2)]\left[ \Lam^2 - m^2
\ln\left(1+\Lam^2/m^2\right)\right]
\label{hcorx}
\eeq
Adopting the parameterization $|\delta \mhs|=|\delta^{(SM)} \mhs +
\delta^{(\vp)} \mhs| = D_t \mhs$ \cite{Barbieri:2006dq}, we can
determine the value of $\lamx$ needed to suppress $\delta \mhs$ to a
desired level ($D_t$) as a function of $m$, for any choice of $\mh$ and
$\Lam$; examples are plotted in fig.\ref{lam} for $N_\vp=6$.
It should be noted that (in contrast to SUSY~\footnote{Note that in
SUSY the corresponding logarithmic stop contributions survive and
constitute a source of concern.}) the logarithmic terms in
(\ref{hcorx}) can be relevant in canceling large contributions to
$\delta \mhs$.   
It is important to note that the required value of $ \lamx $ is smaller for
larger $ \mh $, and can also be reduced increasing the number of 
singlets $N_\vp$. 
When $m \ll \Lam$, the $\lamx$ needed for the amelioration of the
hierarchy problem is insensitive to $m$, $D_t$ or $\Lam$; as
illustrated in fig.\ref{lam}; analytically we find
\beq 
\lamx =N_\vp^{-1}
\left\{4.8 - 3 (\mh/v)^2 + 2D_t
[2\pi/(\Lam/\tev)]^2\right\}
\left[1-m^2/\Lam^2\ln \left(m^2/\Lam^2\right)\right]
+\ocal \left(m^4/\Lam^4\right)\,.
\label{laxaprox}
\eeq 
Since we consider $\lamx \sim 1 $, it is pertinent to estimate the
effects of higher order corrections~\cite{Einhorn:1992um} to 
(\ref{hcor}).  In general, the fine tunning condition reads
($\mh$ was chosen as a renormalization scale):
\beq 
|\delta^{(SM)} \mhs + \delta^{(\vp)} \mhs + \Lam^2
\sum_{n=1}f_n(\lamx, \dots)
\left[ \ln(\Lam/\mh) \right]^n |
= D_t \mhs\,,
\label{hor}
\eeq 
where the coefficients $f_n(\lamx, \dots)$ can be determined
recursively~\cite{Einhorn:1992um}, with the leading
contributions being generated by loops containing powers of $\lamx$:
$f_n(\lamx, \dots)\sim [\lamx/(16 \pi^2)]^{n+1}$. To estimate
these effects consider the case where 
$\delta^{(SM)} \mhs + \delta^{(\vp)} \mhs = 0$ at one loop then,
keeping only
terms $\propto \lamx^2$,  we find, at 2 loops, 
$D_t\simeq [N_\vp \lamx/(16\pi^2)]^2 (\Lam/\mh)^2$. 
Requiring $D_t \lsim 1$ implies $\Lam \lsim 4\pi^2 \mh \simeq 5-8\tev$ for
$\mh=130-210\gev$, respectively. 

It should be emphasized that in the model proposed here
the hierarchy problem is softened (by lifting the cutoff
 to $\sim 8\tev$) only if
$\lamx$, $ \Lam $ and $m$ are appropriately fine-tuned;
this fine tuning, however, is significantly less 
dramatic than in the SM. One can
investigate this issue quantitatively and determine the
range of parameters that corresponds to a given level of fine-tuning
as in~\cite{Kolda:2000wi};
 we will return to this in a future publication~\cite{BJ}.


\paragraph{Dark matter}
The singlets $\vp_i$ also provide a natural dark-matter (DM) candidates
(see \cite{Silveira:1985ke}, \cite{Burgess:2000yq} for the one singlet case).  
Following \cite{Kolb:1990vq} one can easily estimate the amount of the present
DM abundance; we will assume for simplicity that all the $ \vp_i $
are equally abundant (e.g. as in the $O(N_\vp)$ limit). 
The thermal averaged cross-section for  
singlet annihilations into SM final states  $\vp_i \vp_i \to SM~SM$
in the non-relativistic approximation, and for $ m \gg \mh $,
equals
\beq
\left\langle \sigma_i v \right\rangle
\simeq 
 \frac{\lamx^2}{8 \pi m^2}
+ \frac{\lamx^2 v^2 \Gamma_h(2 m)}{8 m^5 }
 \simeq \frac{1.73}{8\pi} \frac{\lamx^2}{m^2}
\label{sigv}
\end{equation}
where the first contribution is from the $hh$ final state 
(keeping
only the s-channel Higgs exchange; the  t and u channels can be neglected
since $m \gg \mh$) while the second contribution is from all other
final states; $\Gamma_h(2m) \simeq 0.48 \tev ( 2m/ 1 \tev)^3$ 
is the Higgs width
calculated when the Higgs boson mass equal $2 m$.

From this the freeze-out temperature $x_f=m/T_f$ is given by
\beq 
x_f=\ln\left[0.038
\mpl m \vevof{\sigma_i v}/(g_\star x_f)^{1/2} \right]
\eeq
where  $g_\star$ counts  relativistic degrees of 
freedom 
at annihilation and $\mpl$ denotes the Planck mass. In the 
range of 
parameters we are interested in, $x_f\sim 12-50$ 
while $m \sim 1-2 \tev$, so that
this is a case of cold dark matter. 
Then the present density of $\vp_i$ is
given by
\beq
\Omega_\vp^{(i)} h^2 = 1.06\cdot 10^9 x_f/(g_\star^{1/2} \mpl \langle
\sigma_i v \rangle \gev)\,.
\label{om}
\eeq

\begin{figure}[ht]
\centering
\includegraphics[width=5cm]{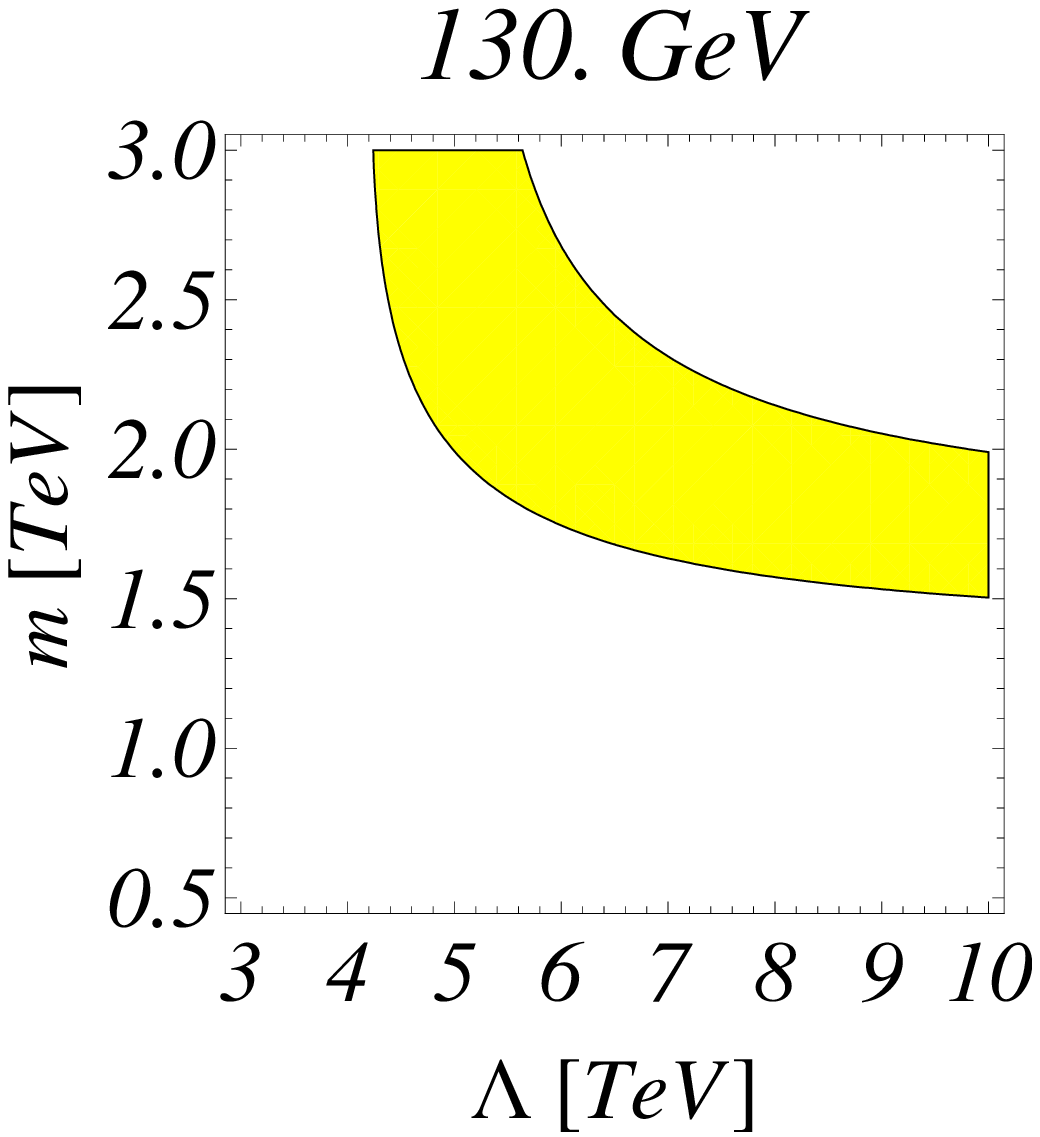}
\includegraphics[width=5cm]{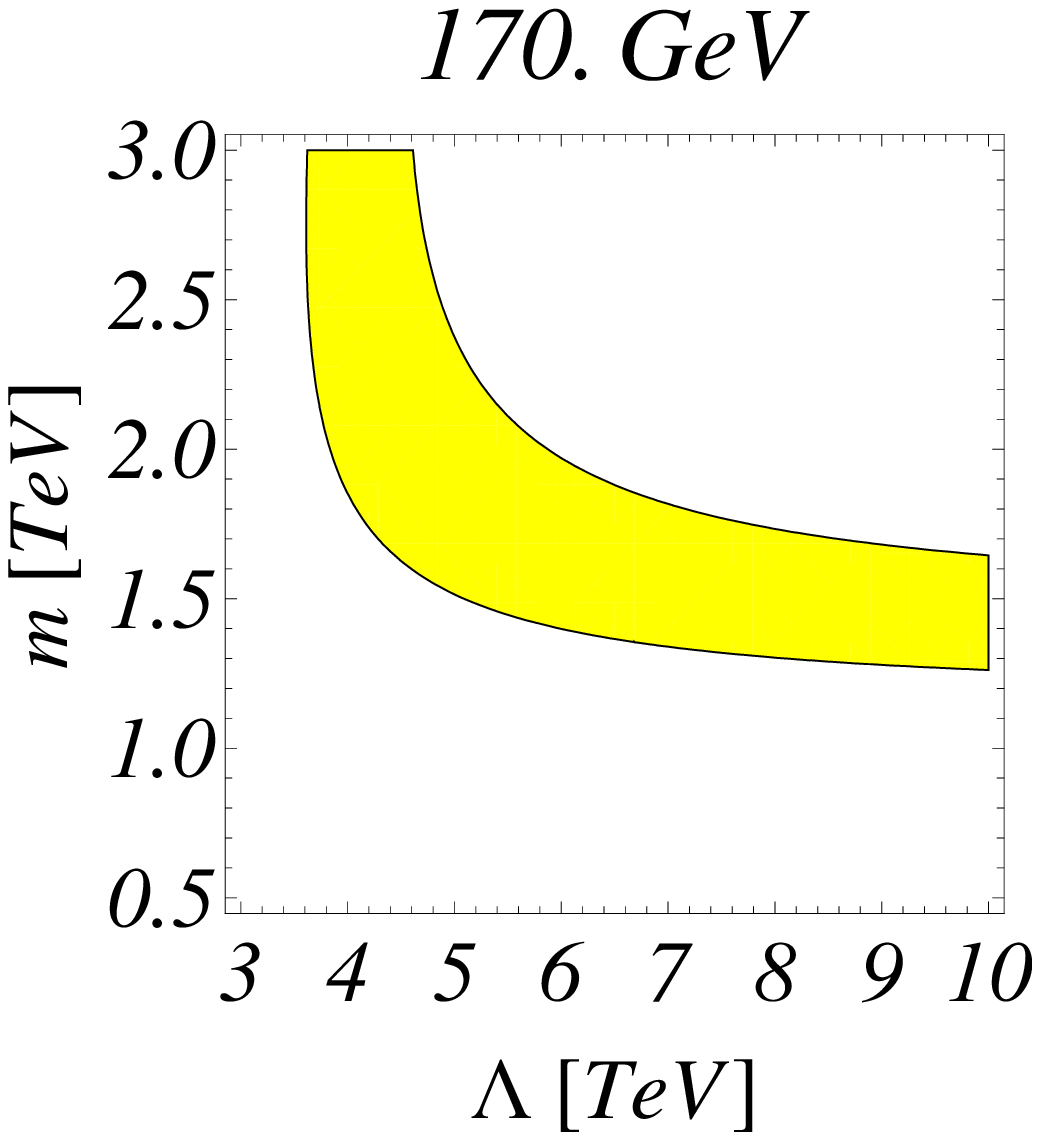}
\includegraphics[width=5cm]{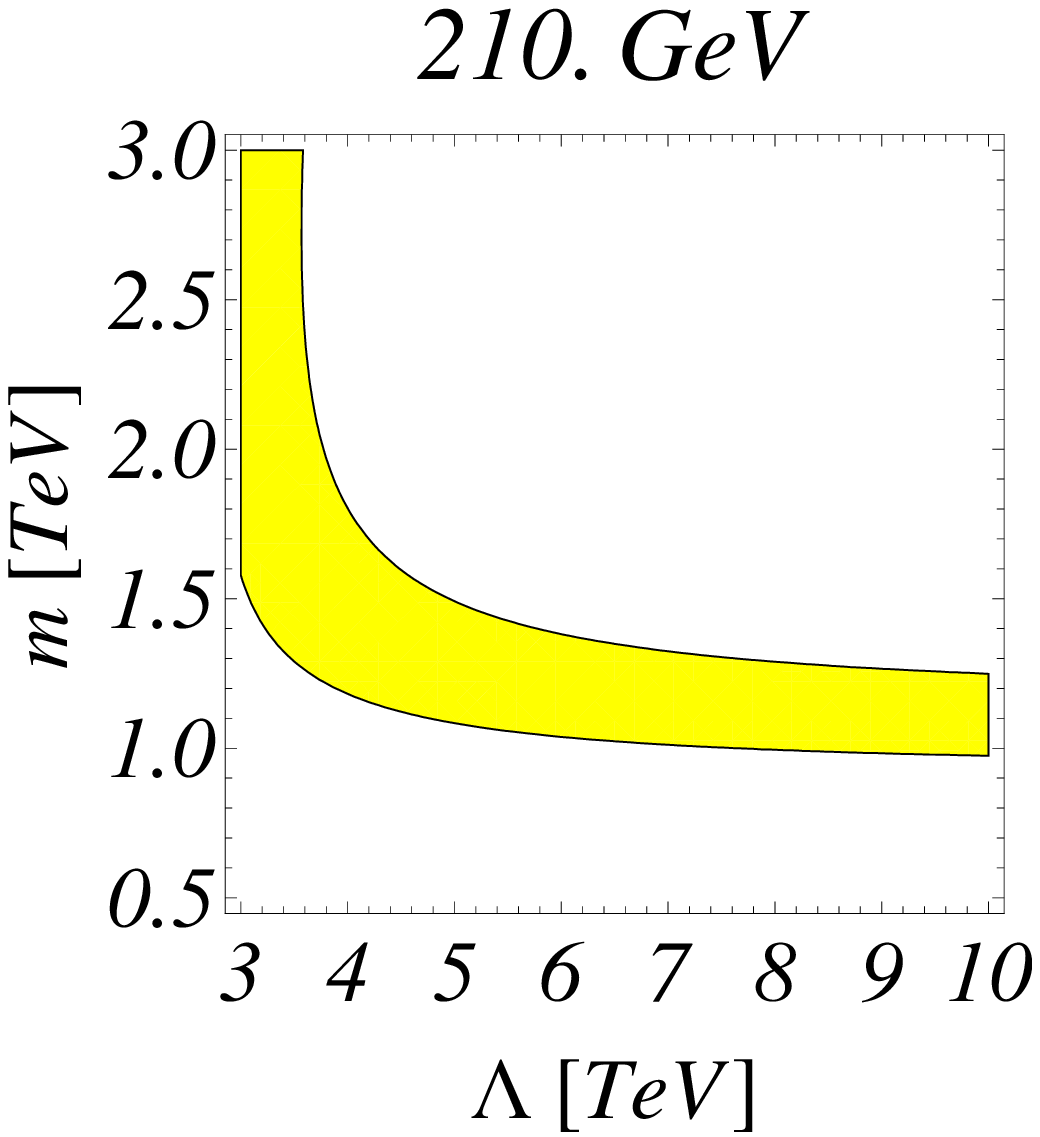}
\caption{The allowed region in the $(m,\Lam)$ plane for $ D_t = 0$, $N_\vp=6$
and $\sum_{i=1}^{N_\vp}\Omega_\vp^{(i)} h^2= 0.106
\pm 0.008$ at the $3\sigma$ level for $\mh=130, 170, 210 \gev$ (as indicated above 
each panel).}
\label{mLamplot}
\end{figure}

Finally, the requirement that the $\vp_i $ account for
the inferred DM abundance,
$\Omega_{DM}h^2=\sum_{i=1}^{N_\vp}\Omega_\vp^{(i)} h^2= 0.106
\pm 0.008$~\cite{Amsler:2008zz},
can be used to fix $\langle \sigma_i v \rangle$, 
which translates into a relation  $\lamx = \lamx(m)$
thorugh the use of
(\ref{sigv}).
Substituting this into $|\delta \mhs|=D_t \mhs$, we find a relation
between $m$ and $\Lam$ (for a given $D_t$), which we plot in
fig.\ref{mLamplot} for $N_\vp=6$. 
It is important to stress that it is possible to find parameters
$\Lam$, $\lamx$ and $m$ such that {\it both} the hierarchy is
ameliorated to the prescribed level {\it and} such that $\Omega_\vp
h^2$ is consistent with the DM requirement (we use a $3\sigma$ interval).
It also is useful to note  that the singlet
mass (as required by the DM) scales with their multiplicity as $N_\vp^{-3/2}$, therefore 
increasing $N_\vp$ implies smaller scalar mass, e.g. changing $N_\vp$ from 1 to 6
leads to the  reduction of mass by a factor $\sim 15$.

%

\paragraph{Neutrinos}

We now discuss consequences of the existence of $\vp$ for the leptonic
sector, which we assume consists of the SM fields plus three
right-handed neutrino fields~\footnote{The arguments presented below
remain essentially the same when a different number of right-handed
neutrinos is present.}  $\nu_{i\,R}$ ($i=1,2,3$) that are also gauge
singlets; in this section we assume only one singlet for simplicity.
The relevant Lagrangian is then
\beq
\lcal_Y = -\bar L Y_l H l_R - \bar L Y_\nu \tilde H \nu_R -
\frac12 \overline{(\nu_R)^c} M \nu_R - \vp \overline{(\nu_R)^c}
Y_\vp \nu_R + \hc
\label{lyuk}
\eeq 
where $L=(\nu_L,l_L)^T$ is a SM lepton isodoublet  and $l_R$ 
a charged lepton
isosinglets (we omit family indices); we will assume that the
see-saw mechanism explains the smallness of three light neutrino
masses, and accordingly we require $M\gg M_D \equiv Y_\nu
v/\sqrt{2}$. The symmetry of the potential under $\vp \to -\vp$ can be
extended to (\ref{lyuk}) by requiring
\beq
L \to S_L L, \; l_R \to S_{l_R} l_R, \; \nu_R \to S_{\nu_R} \nu_R 
\label{trans}
\eeq 
where the unitary matrices $ S_{L, l_R, \nu_R} $ obey
\beq
\quad S_L^\dagger Y_l S_{l_R} = Y_l, \quad \quad S_L^\dagger Y_\nu
S_{\nu_R} = Y_\nu, \quad S_{\nu_R}^T M S_{\nu_R} = + M, \quad
S_{\nu_R}^T Y_\vp S_{\nu_R} = - Y_\vp
\label{inv}
\eeq

In order to determine the consequences of this symmetry we find it
convenient to adopt the basis in which $M$ and $Y_l$ are real and
diagonal; for simplicity we will also assume that $M$ has no
degenerate eigenvalues.  Then the last two conditions in (\ref{inv})
imply that $S_{\nu_R}$ is real and diagonal, so its elements are $\pm
1$.  For $3$ neutrino species there are then two possibilities (up to
permutations of the basis vectors): we eiher have 
$ S_{\nu_R} = \pm \mati,~ Y_\vp =0 $, or, more interestingly,
\beq
S_{\nu_R} = \epsilon \; \hbox{diag}(1,1,-1); \quad
Y_\vp = \bpm{0 & 0 & b_1 \cr 0 & 0 & b_2 \cr b_1 & b_2 & 0 }\epm ,
\qquad \epsilon = \pm1\,,
\label{snur.sol}
\eeq
where $ b_{1,2} $ are, in general,
complex. The first  conditions in (\ref{inv}) now requires
$S_{l_R}=S_L$
with
\beq
S_L = \hbox{diag}(s_1, s_2,s_3) , \quad |s_i| = 1 
\label{SL}
\eeq
Before discussing the explicit solutions for $Y_\nu$, we first
diagonalize (to leading order in $M^{-1}$) the neutrino mass matrix
in terms of the light ($n$) and heavy ($N$) eigenstates:
\beq
\lcal_m=-(\bar n M_n n + \bar N M N/2) \quad {\rm with } \quad
 M_n = \mu^* P_R + \mu P_L, ~~ \mu = 
- 4 M_D M^{-1} M_D^T
\eeq
where $n$ and $N$ are related to $\nu_R$ and $\nu_L$ through $\nu_L =
n_L + (M_D M^{-1}) N_L$ and $\nu_R = N_R - (M^{-1} M_D^T)
n_R$. 

The remaining condition in (\ref{inv}) allows ten (up to permutations
of the basis vectors) inequivalent solutions 
for $Y_\nu$~\footnote{ The conditions (\ref{inv}) where
also investigated in \cite{Low:2005yc}.}.  
Of those, assuming single massless neutrino and the 
absence of $\vp\to n_i n_j$ decays, only one is acceptable;
it  corresponds to $ s_{1,2,3} = \epsilon $ (cf. (\ref{snur.sol})).
To compare our results with the data, we use the so-called 
tri-bimaximal~\cite{Harrison:2002er} lepton mixing matrix that 
corresponds to $\theta_{13}=0$, $\theta_{23}=\pi/4$ 
and $\theta_{12}=\arcsin(1/\sqrt{3})$.
One can undo the diagonalization of light neutrino mass matrix and check 
against the one implied by $Y_\nu$ as a consequence of (\ref{inv}).
We find that 
there are only two possible forms of $Y_\nu$ 
that are consistent with (\ref{inv}) and independent of $M$, and
that agree with tri-bimaximal mixing:
\beq
Y_\nu=\left(\baa{ccc}a&b&0\\-a/2&b&0\\-a/2&b&0\eaa\right)\,,
\baa{l}m_1=-3v^2a^2/M_1\\m_2=-6v^2b^2/M_2\\m_3=0\eaa
\;\;\hbox{and}\;\;
Y_\nu=\left(\baa{ccc}a&b&0\\a&-b/2&0\\a&-b/2&0\eaa\right)
\baa{l}m_1=-3v^2b^2/M_2\\m_2=-6v^2a^2/M_1\\m_3=0\eaa
\label{ynu}
\eeq
where $a$ and $b$ are real (for simplicity) parameters.
The resulting mass spectrum is
consistent with the observed pattern of neutrino mass differences, see
e.g. \cite{Altarelli:2007gb}.
For this solution only $N_3$ and $ \vp $
are odd under the $Z_2$ symmetry hence the  $\vp$ will
be absolutely stable if $m<M_3$.

It is noteworthy that the presence of $ Y_\vp $ also leads to an
additional contribution $-(\Lam/\pi)^2 \hbox{tr} Y_\vp^2$ to $\delta
m^2$ (we assumed $Y_\vp$ real for simplicity) so the neutrinos can be
used to ameliorate the little hierarchy problem associated with $ m $
(for this however $Y_\vp$ cannot be too small) thereby
``closing'' the solution to the little hierarchy problem in a spirit
similar to supersymmetry. This interesting scenario will be discussed
elsewhere~\cite{BJ}.

\paragraph{Conclusions}
We have shown that the addition of real scalar singlets $\vp_i$ to the SM may
ameliorate the little hierarchy problem (by lifting the cutoff $\Lam$
to multi TeV range) and also provide realistic candidates for DM. In
the presence of right-handed neutrinos this scenario allows 
a light neutrino mass matrix texture that is consistent
with experimental data while preserving all the
successes of leptogenesis as an explanation for the baryon
asymmetry.

\acknowledgments

This work is supported in part by the Ministry of Science and Higher
Education (Poland) as research project N~N202~006334 (2008-11) 
and by the U.S. Department of Energy
grant No.~DE-FG03-94ER40837.  B.G. acknowledges support of the
European Community within the Marie Curie Research \& Training
Networks: "HEPTOOLS" (MRTN-CT-2006-035505), and "UniverseNet"
(MRTN-CT-2006-035863), and through the Marie Curie Host Fellowships
for the Transfer of Knowledge Project MTKD-CT-2005-029466.
B.G. thanks S.~Davidson, S.~Pokorski and Z.~Lalak for useful conversations
and M.~Pospelov and J.~McDonald for correspondence concerning the 
$\vp$ annihilation cross section. JW would like to thank E. Ma for
helpful conversations.

\end{document}